\documentclass[12pt,epsf]{article}
\usepackage{graphicx}
\usepackage{cite}
\usepackage[hidelinks=true]{hyperref}

\begin{document}

\def\a{\alpha}
\def\b{\beta}
\def\c{\varepsilon}
\def\d{\delta}
\def\e{\epsilon}
\def\f{\mathcal{A}}
\def\g{\gamma}
\def\h{\theta}
\def\k{\kappa}
\def\l{\lambda}
\def\m{\mu}
\def\n{\nu}
\def\p{\psi}
\def\q{\partial}
\def\r{\rho}
\def\s{\sigma}
\def\t{\tau}
\def\u{\upsilon}
\def\v{\varphi}
\def\w{\omega}
\def\x{\xi}
\def\y{\eta}
\def\z{\zeta}
\def\D{\Delta}
\def\G{\Gamma}
\def\H{\Theta}
\def\L{\Lambda}
\def\F{\mathcal{A}}
\def\P{\Psi}
\def\S{\Sigma}

\def\o{\over}
\def\beq{\begin{eqnarray}}
\def\eeq{\end{eqnarray}}
\newcommand{\gsim}{ \mathop{}_{\textstyle \sim}^{\textstyle >} }
\newcommand{\lsim}{ \mathop{}_{\textstyle \sim}^{\textstyle <} }
\newcommand{\vev}[1]{ \left\langle {#1} \right\rangle }
\newcommand{\bra}[1]{ \langle {#1} | }
\newcommand{\ket}[1]{ | {#1} \rangle }
\newcommand{\EV}{ {\rm eV} }
\newcommand{\KEV}{ {\rm keV} }
\newcommand{\MEV}{ {\rm MeV} }
\newcommand{\GEV}{ {\rm GeV} }
\newcommand{\TEV}{ {\rm TeV} }
\def\diag{\mathop{\rm diag}\nolimits}
\def\Spin{\mathop{\rm Spin}}
\def\SO{\mathop{\rm SO}}
\def\O{\mathop{\rm O}}
\def\SU{\mathop{\rm SU}}
\def\U{\mathop{\rm U}}
\def\Sp{\mathop{\rm Sp}}
\def\SL{\mathop{\rm SL}}
\def\tr{\mathop{\rm tr}}
\def\mpl{M_{\rm Pl}}

\def\IJMP{Int.~J.~Mod.~Phys. }
\def\MPL{Mod.~Phys.~Lett. }
\def\NP{Nucl.~Phys. }
\def\PL{Phys.~Lett. }
\def\PR{Phys.~Rev. }
\def\PRL{Phys.~Rev.~Lett. }
\def\PTP{Prog.~Theor.~Phys. }
\def\ZP{Z.~Phys. }


\baselineskip 0.7cm

\begin{titlepage}


\vskip 1.35cm
\begin{center}
{\large \bf
 Axion induced SUSY breaking and \\
 focus point gaugino mediation
}
\vskip 1.2cm
Keisuke Harigaya$^{1,2,3}$, Tsutomu T. Yanagida$^{4,5,6}$ and Norimi Yokozaki$^{7,8}$
\vskip 0.4cm

{\it  
$^{1}$ Theoretical Physics Department, CERN, Geneva, Switzerland \\
$^{2}$Department of Physics, University of California, Berkeley, California 94720, USA \\
$^{3}$Theoretical Physics Group, Lawrence Berkeley National Laboratory, Berkeley, California 94720, USA\\
     $^{4}$ Tsung-Dao Lee Institute, Shanghai Jiao Tong University, Shanghai 200240, China \\
     $^{5}$Kavli IPMU (WPI), UTIAS, University of Tokyo, Kashiwa, Chiba 277-8583, Japan\\
     $^{6}$Hamamatsu Professor\\
     $^{7}$Zhejiang Institute of Modern Physics and Department of Physics, Zhejiang University,
Hangzhou, Zhejiang 310027, China \\
     $^{8}$Department of Physics, Tohoku University, Sendai, Miyagi 980-8578, Japan}

\vskip 1.5cm

\abstract{
We consider a scenario where supersymmetry breaking, its mediation, and the cancellation of the theta parameter of $SU(3)_C$ are all caused by a single chiral multiplet. The string axion multiplet is a natural candidate of such a superfield. We show that the scenario provides a convincing basis of focus point gaugino mediation, where the electroweak scale is explained with a moderate tuning among the parameters of the theory.
}

\end{center}
\end{titlepage}

\setcounter{page}{2}

\section{Introduction}
An axion with a large decay constant
between the unification scale and the Planck scale
is one of interesting predictions in string theory~\cite{Svrcek:2006yi}, which solves the strong CP problem in QCD~\cite{Peccei:1977hh,Peccei:1977ur,Weinberg:1977ma,Wilczek:1977pj}, and is a good candidate for the 
dark matter (DM) observed today.
In this paper, we consider a framework where the string axion causes a spontaneous supersymmetry (SUSY) breaking under the condition of the vanishing cosmological constant, via the mechanism of the gravitational SUSY breaking~\cite{Izawa:2010ym}.

The framework provides a convincing basis of focus point gaugino mediation~\cite{Yanagida:2013ah}~\footnote{We refer readers to~\cite{Feng:1999mn,Feng:1999zg} for an original proposal for a focus point scenario where scalar masses are much larger than gaugino masses, and to~\cite{Feng:2012jfa,Brummer:2013dya,Harigaya:2015iva,Harigaya:2015jba} for recent discussion on focus point scenarios.} 
with vanishing soft masses for sfermions at a high energy scale. The electroweak symmetry breaking (EWSB) scale is explained with $O(1)$\% tuning between the gaugino mass and the SUSY invariant mass of the Higgs multiplet. 
This is highly non-trivial since there are severe lower bounds on the masses of SUSY particles from the Large Hadron Collider (LHC) \cite{Buckley:2016kvr,CMS:2019zmd, ATLAS:2020syg,ATLAS:2021hza} as well as the observed Higgs mass of 125 GeV~\cite{Aad:2012tfa,Chatrchyan:2012xdj}, which requires large stop masses~\cite{Okada:1990gg,Okada:1990vk,Ellis:1990nz,Haber:1990aw}.
The mild tuning to explain the EWSB scale is achieved with relatively large bino and wino masses compared to a gluino mass at a high energy scale.%
\footnote{
The importance of the non-universal gaugino masses to reduce the fine-tuning has been noticed in Refs.~\cite{Kane:1998im, BasteroGil:1999gu, Abe:2007kf, Martin:2007gf, Horton:2009ed, Younkin:2012ui, Gogoladze:2012yf}.} 
In our setup, the mass ratios of the gauginos are fixed at the Planck or string scale by the anomaly coefficients of the shift symmetry, i.e., integer numbers.

The framework would be also attractive from the view point of minimality; a single chiral multiplet, an axion multiplet, is responsible for SUSY breaking, its mediation to the standard model sector at a high energy scale with focus point gaugino mediation, dark matter, and a solution to the strong CP problem.

This paper is organized as follows. In section \ref{sec:SB}, we briefly review the axion-induced SUSY breaking. In section \ref{sec:fpgm}, focus point gaugino mediation and its LHC signals are discussed. We also show that the light Higgsino can be detected at future direct-detection experiments. In section \ref{sec:cosmology}, the cosmological aspects of our model, especially an imprint on the cosmic microwave background, are discussed. 
Finally section \ref{sec:discussion} is devoted to the conclusion and discussion.

\section{Axion induced SUSY breaking}
\label{sec:SB}
In this section we briefly review  the mechanism of SUSY breaking by a string axion multiplet $\mathcal{A}$.
We will see that SUSY is necessarily broken by the F-term of $\mathcal{A}$ when the cosmological constant vanishes.
The string axion multiplet $\mathcal{A}$ enjoys a shift symmetry $\mathcal{A} \to \mathcal{A} + i {\mathcal R}$, where $\mathcal{R}$ is a real constant.
A K{\"a}hler potential $K$ and a super potential $W$ consistent with the shift symmetry are given by
\begin{eqnarray}
\label{eq:KW}
K = K(\mathcal{A} + \mathcal{A}^\dag) \equiv K(x), \ \ W = \mathcal{C}.
\end{eqnarray}
Note that the superpotential is independent of $\mathcal{A}$ due to the shift symmetry. Here, we assume the constant $\mathcal{C} \neq 0$.

The scalar potential of $\mathcal{A}$ is given by
\begin{eqnarray}
V = e^{K} \left[
\left(\frac{\partial K}{\partial x}\right)^2 \left(\frac{\partial^2 K}{\partial x^2}\right)^{-1}-3
\right]  |\mathcal{C}|^2,
\end{eqnarray}
where we take the units of the reduced Planck mass $\mpl=1$. The condition of the vanishing cosmological constant, $V=0$, is satisfied if 
\begin{eqnarray}
\left.
\left(\frac{\partial K}{\partial x}\right)^2 \left(\frac{\partial^2 K}{\partial x^2}\right)^{-1}
\right|_{x=\left<x\right>}=3, \label{eq:vcc}
\end{eqnarray}
where $\left<x\right>$ is a vacuum expectation value determined by the stationary condition,
\begin{eqnarray}
\left.\frac{\partial }{\partial x} 
\left[
\left(\frac{\partial K}{\partial x}\right)^2 \left(\frac{\partial^2 K}{\partial x^2}\right)^{-1}
\right] \right|_{x=\left<x\right>}= 0.
\label{eq:vst}
\end{eqnarray}
The F-term of $\mathcal{A}$ is then given by
\begin{eqnarray}
\left.
F_\mathcal{A} = -e^{K/2} \left(\frac{\partial K}{\partial x}\right)
\left(\frac{\partial^2 K}{\partial x^2}\right)^{-1} \right|_{x=\left<x\right>} \mathcal{C}^*,
\end{eqnarray}
which is non-zero as long as Eq.~(\ref{eq:vcc}) is satisfied, i.e., the cosmological constant vanishes~\cite{Izawa:2010ym}.
Notice that the argument of $F_\mathcal{A}$ is aligned to that of $\mathcal{C}^*$ since $K$ is a real function of $x$. 
This alignment is an important feature of our framework; the dangerous SUSY CP problem is absent, as discussed in section \ref{sec:fpgm}.

\section{Focus point gaugino mediation}
\label{sec:fpgm}
Gaugino mediation was proposed to suppress the flavor-changing neutral currents
taking the sequestered K{\"a}hler potential~\cite{Inoue:1991rk,Kaplan:1999ac,Chacko:1999mi}. 
Explicitly, we consider 
\begin{eqnarray}
	K = -3 \mpl^2 \ln \left[1 - \frac{M_*^2 g(\mathcal{A} + \mathcal{A}^\dag)}{3 \mpl^2}  - \frac{f_{\rm vis}}{3 \mpl^2} \right],
\end{eqnarray}
where $M_*$ is a cut-off scale around $M_{\rm Pl}$ and $f_{\rm vis}$ is a function of quark, lepton, gauge, and Higgs multiplets. 
With the sequestered K{\"a}hler potential
all soft SUSY breaking masses beside the gaugino masses vanish at a high energy scale.%
\footnote{As is shown in~\cite{Kugo:2010fs,Harigaya:2015iva}, the vanishing soft mass may be also understood by a Nambu-Goldstone nature of chiral multiplets~\cite{Buchmuller:1982xn, Buchmuller:1982tf,Goto:1990me}.}
Therefore, the gaugino masses are only parameters of the SUSY breaking, determining the low-energy mass spectrum of the SUSY particles.
In the axion-induced SUSY breaking scenario the gaugino masses are given by the couplings of the (caninically normalized) axion multiplet $\mathcal{A}_c$ to the gauge multiplets.%
\footnote{Mediation of SUSY breaking by an axion multiplet without a focus point is discussed in~\cite{Higaki:2011bz,Baryakhtar:2013wy,Harigaya:2017dgd}.}
The couplings are fixed by the anomaly indices of the shift symmetry. The relevant part of the Lagrangian is 
\begin{eqnarray}
\mathcal{L} \supset \frac{\sqrt{2}}{32\pi^2 f_a} \int d^2 \theta \, \mathcal{A}_c \left[ k_1 \mathcal{W}_1^2
+ k_2 \mathcal{W}_2^2
+ k_3 \mathcal{W}_3^2 \right], \label{eq:aww}
\end{eqnarray}
where $k_1$, $k_2$, and $k_3$ are integers corresponding to the anomaly indices of the shift-symmetry;\footnote{Here we assume the quantization of the $U(1)_Y$ charge, which is the case for $U(1)$ gauge theories in low energy effective theories of the string theory, embedded into non-abelian gauge symmetries, embedded into the diffeomorphism of higher dimensional theories, or with a Dirac monopole in the spectrum.	The quantization is also supported by the argument from the absence of exact global symmetries~\cite{Banks:2010zn}.} 
$\mathcal{W}_1$, $\mathcal{W}_2$, and $\mathcal{W}_3$ are field strength superfields of $U(1)_Y$, $SU(2)_L$, and $SU(3)_C$, respectively; $f_a$ is the decay constant of the string axion, which is $\sim M_*^2/\mpl$ (see Appendix A).
The coupling with $\mathcal{W}_3$ is responsible for the mass of the axion, and hence the solution to the strong CP problem.
The integers $k_1, k_2$ and $k_3$ can arise from string theory (see e.g.,~\cite{Brignole:1993dj}, although the SUSY breaking field there is not identified with the QCD axion and the sequestered K{\"a}hler potential is not constructed). They can also arise in four-dimensional field theory from~\cite{Higaki:2011bz},  
\begin{eqnarray}
	W = M' e^{-q_a \mathcal{A}} \Psi_{a}^{I_a} \bar{\Psi}_{a}^{I_a},
\end{eqnarray}
where $a$ represents a gauge index, $I_a=1 \dots N_a$ is the number of charged matters, and $\Psi_{a}^{I_a}$ are assumed to be fundamental representations of the gauge groups. $\Psi_{a}^{I_a} \bar{\Psi}_{a}^{I_a}$ transforms as $\Psi_{a}^{I_a} \bar{\Psi}_{a}^{I_a} \to \exp[i q_a \mathcal{R}] \Psi_{a}^{I_a} \bar{\Psi}_{a}^{I_a}$ under the shift symmetry.
In this case, $k_a = q_a N_a$ (the charge $q_a$ is assumed to be quantized). The gaugino masses read 
\begin{eqnarray}
	M_a = \frac{k_a  g_a^2}{16\pi^2}  \left( \frac{\sqrt{6} \mpl}{f_a} \right) m_{3/2}.
\end{eqnarray}
It should be stressed that the ratios of the gaugino masses are fixed by the anomaly indices of the shift symmetry. This feature leads to the scenario of focus point gaugino mediation, where the EWSB is relatively insensitive to the masses of the SUSY particles.
In the following we explain how the focus point behavior is achieved.

The EWSB scale is determined by the stationary conditions:
\begin{eqnarray}
\frac{g_Y^2 + g_2^2}{4} v^2 &\simeq& \left[ -\mu^2 
- \frac{(m_{H_u}^2  + \frac{1}{2 v_u}\frac{\partial \Delta V}{\partial v_u} ) \tan^2\beta}{\tan^2\beta-1} 
 + \, \frac{m_{H_d}^2 + \frac{1}{2 v_d}\frac{\partial \Delta V}{\partial v_d} }{\tan^2\beta-1} \right]_{M_{\rm stop}}, \nonumber \\
\frac{B_\mu \,(\tan^2\beta+1)}{\tan\beta} &\simeq& \left[ m_{H_u}^2 +\frac{1}{2 v_u}\frac{\partial \Delta V}{\partial v_u} + m_{H_d}^2  + \frac{1}{2 v_d}\frac{\partial \Delta V}{\partial v_d} + 2\mu^2 \right]_{M_{\rm stop}}, \label{eq:ewsb}
\end{eqnarray}
wheree $g_Y$ and $g_2$ are gauge coupling constants of $U(1)_Y$ and $SU(2)_L$ respectively; $v \,(\equiv \sqrt{v_u^2 + v_d^2})$ is the EWSB scale and $\tan\beta\,(\equiv v_u/v_d)$ is a ratio of vacuum expectation values of $H_u$ and $H_d$;
$\mu$ is the SUSY invariant Higgsino mass term; $m_{H_u}^2$ and $m_{H_d}^2$ are soft SUSY breaking masses for $H_u$ and $H_d$, respectively; $\Delta V$ is one-loop contributions to the Higgs potential. The above stationary conditions are evaluated at the stop mass scale $M_{\rm stop}$.

For a large value of $\tan\beta$, the EWSB scale is dominantly determined by $m_{H_u}^2$ and $\mu^2$.
The soft SUSY breaking mass for the up-type Higgs at the stop mass scale can be written as
\begin{eqnarray}
m_{H_u}^2 (4\,{\rm TeV}) &\simeq& 0.012 M_1^2 + 0.246 M_2^2  -1.025 M_3^2 \nonumber \\
&-& 0.004 M_1 M_2 -0.113 M_2 M_3 -0.017 M_1 M_3,
\end{eqnarray}
for $\tan\beta=30$, $m_t=173.34$\,GeV, and $\alpha_s(m_Z)=0.1181$. Here, $M_1$, $M_2$, and $M_3$ are bino, wino, and gluino mass at the scale $M_{\rm in}=2\times 10^{16}$\,GeV, respectively. We see that, for instance, when $M_1:M_2:M_3=k_1:k_2:k_3 = 6:2:1$, $m_{H_u}^2$ becomes significantly smaller than the gluino mass scale.\footnote{%
If the $R$ charge of $H_u H_d$ are zero and the matter fields are unified into ${\bf 10}$ and ${\bf 5}^*$ representations of $SU(5)$, the anomaly coefficients of $Z_{N,R}$\,-\,$SU(3)^2$ and $Z_{N,R}$\,-\,$SU(2)^2$ are 6 and 2 (mod $N$), respectively~\cite{Kurosawa:2001iq}. The desired ratio, $M_3 : M_2 = 1 : 2$, may be explain by further assuming $Z_{10,R}$ and the anomaly cancellations through the Green-Schwartz mechanism~\cite{Green:1984sg} with the shift of $Z$. (See discussion in Ref.~\cite{Yanagida:2013uka}.)  The $\mu$-term can be written as $\mu=c' m_{3/2}$ in this case. 
In the case where three pairs of ${\bf 5} + {\bf 5}^*$ are introduced and the $R$ charge of ${\bf 5}{\bf 5}^*$ is 2, 
the requirement of the non-anomalous $Z_{8,R}$ may explain $M_3: M_2 = 1:3$. 
} 

The required ratio is different from unity and cannot be embedded into a simple $SU(5)$ unification.
The ratio $k_1\neq k_2 \neq k_3$ is consistently obtained, for instance, in the framework of product group unification~\cite{Yanagida:1994vq,Hotta:1995cd} as shown in~\cite{ArkaniHamed:1996jq} (see also \cite{Harigaya:2015jba}).
In product group unification, the unification of quarks and leptons into $SU(5)$ multiplets is maintained, and the gauge coupling unification is predicted if gauge couplings other than that of $SU(5)$ is large at the symmetry breaking scale. It should be noted that four dimensional $SU(5)$ unification theories necessarily suffer from the doublet-triplet splitting problem and a too large R symmetry breaking scale~\cite{Goodman:1985bw,Witten:2001bf,Harigaya:2015zea}, while product group unification does not.

\begin{figure}[!t]
\begin{center}
\includegraphics[scale=1.15]{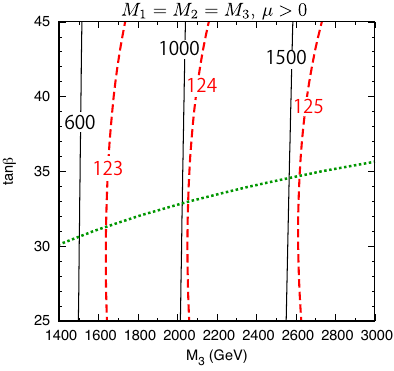}
\includegraphics[scale=1.15]{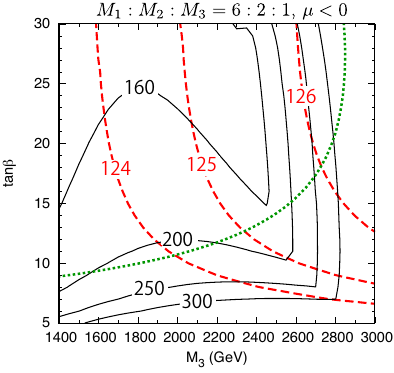}
\caption{The contours of $\Delta$ (black solid line) and $m_h$ (red dashed line). 
In the left (right) panel, $M_1=M_2=M_3$ $(M_1:M_2:M_3=6:2:1)$. 
On the green dotted lines, $B_\mu$-term vanishes at $M_{\rm in}$.
Here, $\alpha_s(m_Z)=0.1181$ and $m_t = 173.34$\,GeV.
}
\label{fig:delta1}
\end{center}
\end{figure}

\begin{table*}[!t]
\caption{\small Mass spectra in sample points. At the point {\bf II}, three pairs of ${\bf 5}+\bar {\bf 5}$ are introduced at the scale $M_5$. Here, $\tan\beta$ is determined to satisfy $B_\mu (M_{\rm in})=0$.
}
\label{tab:sample}
\begin{center}
\begin{tabular}{|c||c|c|c|c|}
\hline
Parameters & Point {\bf I} & Point {\bf II}   \\
\hline
$M_3$ (GeV) & 2500  & 2600    \\
$M_1/M_3$  & 6  & 5    \\
$M_2/M_3$  & 2 & 3  \\
$M_5$ (GeV) & -  & $10^4$   \\
\hline
Particles & Mass (GeV) & Mass (GeV)  \\
\hline
$\tilde{g}$ & 5250 & 2380  \\
$\tilde{q}$ & 4730\,-\,5420 & 3300\,-\,4910  \\
$\tilde{t_{1,2}}$ & 4490,\, 4830 & 1750,\,4110  \\
$\tilde{\chi}_{1}^\pm$ & 837 & 576   \\
$\tilde{\chi}_{2}^\pm$ & 4100 & 2880   \\
$\tilde \chi_1^0$         & 835 & 575   \\
$\tilde{\chi}_2^0$        & 837 & 577   \\
$\tilde \chi_3^0$         & 4100 & 2750  \\
$\tilde{\chi}_4^0$       & 6630 & 2880   \\
$\tilde{e}_{L, R}$       & 4190,\,5520 & 4410,\,4060  \\
$\tilde{\tau}_{1,2}$     & 4150,\,5460&  3940,\,4350  \\
$H^\pm$ & 4120 & 4180 \\
$h_{\rm SM\mathchar`-like}$ & 125.4 &  126.3  \\
\hline
$\mu$ (GeV)  &-814   & -564  \\
$\tan\beta$  & 14.2  & 18.6   \\
$\Delta$  & 171  & 137   \\
\hline
\end{tabular}
\end{center}
\end{table*}

In Fig.~\ref{fig:delta1}, the contours of the fine-tuning measure $\Delta$ (black solid lines) and the Higgs boson mass $m_h$ (red dashed lines) are shown, where $\Delta$ is defined by~\cite{Ellis:1986yg, Barbieri:1987fn}
\begin{eqnarray}
\Delta = {\rm max} \left( \left|\frac{\partial \ln v}{\partial \ln M_3}\right| , 
\left|\frac{\partial \ln v}{\partial \ln |\mu|}\right| 
\right).
\end{eqnarray}
The Higgs boson mass $m_h$ is computed using {\tt FeynHiggs 2.13.0}~\cite{Heinemeyer:1998yj, Heinemeyer:1998np, Degrassi:2002fi, Frank:2006yh, Hahn:2013ria,Bahl:2016brp, Bahl:2017aev, Bahl:2018qog} and mass spectra of SUSY particles are evaluated using {\tt SOFTSUSY 4.0.3}~\cite{Allanach:2001kg}. In the left panel, we take $M_1=M_2=M_3$ as in usual gaugino mediation, while in the right panel we take $M_1:M_2:M_3=6:2:1$ at $M_{\rm in}$. On the green dotted lines, $B_\mu$-term vanishes at $M_{\rm in}$: corresponding $\tan\beta$ is a prediction rather than a free parameter. 
In the case with $M_1=M_2=M_3$, $\Delta > 1500$ for $m_h=125$\,GeV. On the other hand, $\Delta \approx 170$ in the case of $M_1:M_2:M_3=6:2:1$ (focus point gaugino mediation). We see that $\Delta$ is significantly reduced in focus point gaugino mediation.

A mass spectrum and $\Delta$ of a sample point are shown in Table.~\ref{tab:sample}. At the point {\bf I}, the spectrum is evaluated in the minimal supersymmetric standard model (MSSM). The squarks and gluino are $\sim5$\,TeV while the Higgsino is light compared to the gluino mass and squark masses.

Further reduction of $\Delta$ is possible if there are extra matter multiplets at an intermediate mass scale~\cite{Moroi:2012kg, Moroi:2016ztz}.
This is because the trilinear coupling among the Higgs and the stops is enhanced due to larger gauge coupling constants at higher energy scales,
and hence the required stop mass to explain the Higgs mass is reduced.
At the point {\bf II}, we introduce three pairs of ${\bf 5}+\bar{\bf 5}$ of $SU(5)$ at $M_5=10^4$\,GeV. 
The mass spectrum of the SUSY particles is computed using {\tt SuSpect 2.4.3}~\cite{Djouadi:2002ze} with a modification of two-loop level renormalization group equations including effects from the vector-like matters. 
Here, $\Delta$ can be as smalle as $\Delta=137$. Also, the gluino mass (and squark masses) can be significantly smaller for $m_h \simeq 125$\,GeV, which can be tested in the future LHC experiment.  At both points in the table, the mass of the lightest CP-even Higgs is calculated by {\tt FeynHiggs}.

\begin{figure}[!t]
\begin{center}
\includegraphics[scale=0.7]{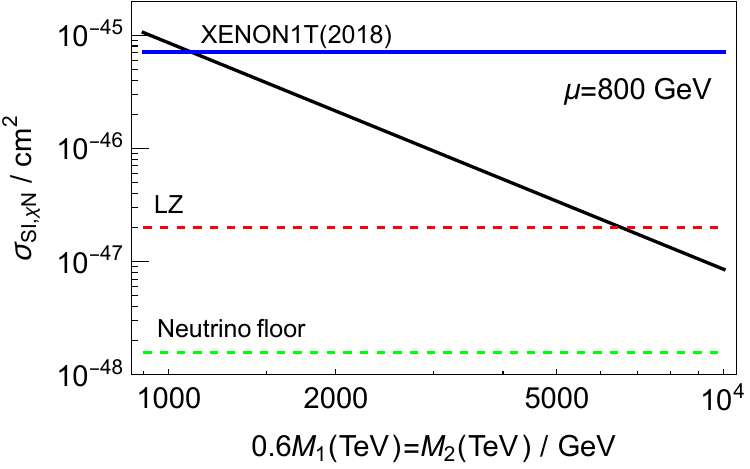}
\caption{The spin-independent LSP-nucleon scattering cross section as a function of $0.6 M_{1}({\rm TeV}) = M_{2}({\rm TeV})$ for $\mu = 800$ GeV.
}
\label{fig:DD}
\end{center}
\end{figure}

In our set up the higgsino-like neutralino is the lightest supersymmetric particle (LSP). With an $R$ parity conservation, the LSP composes a part of dark matter in the universe. The LSP dark matter interacts with nuclei via the Higgs exchange,
\begin{eqnarray}
{\cal L} \simeq \frac{v}{2 \sqrt{2}} \left(
\frac{g_2^2}{M_{2} ({\rm TeV})} + \frac{g_Y^2}{M_1 ({\rm TeV})}
\right) h \tilde{\chi}_1^0 \tilde{\chi}_1^0,
\end{eqnarray}
where we assume a large-${\rm tan}\beta$ limit.
In Fig.~\ref{fig:DD}, the spin-independent LSP-nucleon scattering cross section is shown as a function of $0.6 M_{1}({\rm TeV}) = M_{2}({\rm TeV})$ for $\mu = 800$ GeV.
The constraint from XENON 1T (2018)~\cite{XENON:2018voc}, the future prospect of LZ~\cite{Akerib:2015cja}, and so-called the neutrino floor~\cite{Billard:2013qya} are also shown.
It can be seen that future experiments can cover the parameter space of our model with $\Delta \sim 100$.

We comment on the effect of anomaly mediation~\cite{Randall:1998uk,Giudice:1998xp}.
If $f_a =O(\mpl)$, the gravitino mass $m_{3/2}$ is as large as $O(100)$ TeV and
anomaly mediation generically generates the soft masses of $O(1)$ TeV, which ruins focus point gaugino mediation. However, in our setup with the SUSY breaking by the K{\"a}hler and super potential in Eq.~(\ref{eq:KW}), the vacuum expectation value of the scalar auxiliary component of the supergravity multiplet vanishes~\cite{Izawa:2010ym}. 
Thus unavoidable anomaly mediation determined by the super-diffeomorphism invariance~\cite{Harigaya:2014sfa} also vanishes. (A too large $B_\mu$ term of the Higgs multiplet is also avoided.)
Anomaly mediation caused by the couplings between the F term of the SUSY breaking field and the MSSM fields from the K{\"a}hler potential~\cite{DEramo:2012vvz} also vanishes as we assume the sequestering. 
The remaining possibility is the mediation effect caused by the couplings between the SUSY breaking field and the MSSM fields in path-integral measures. We assume that the path-integral measures of the SUSY breaking field and the MSSM fields are also sequestered from each other. 
The focus point behavior is then not disturbed by anomaly mediation in our setup even if $m_{3/2}= O(1)$ TeV.

Finally, we note that the CP violation in the MSSM sector vanishes~\cite{Iwamoto:2014ywa}.
The sequestering ensures that CP phases from sfermion masses vanish.
Due to the shift symmetry of $\mathcal{A}$, the phases of the gaugino masses are aligned with each others, and hence an $R$ rotation that makes the constant ${\cal C}$ real ensures that all of the gaugino masses are real.
The $\mu$ term can be made real by a PQ rotation.
The $B_\mu$ term is radiatively generated from the gaugino masses, and is also real.%
\footnote{We may introduce a coupling between the down-type Higgs $H_d$ and the SUSY breaking field $\mathcal{A}$ in the K{\"a}hler potential without spoiling the focus point behavior while making tan$\beta$ a free parameter because of a non-zero $B_\mu$ term at a mediation scale. Even in this case the shift symmetry ensures that the coupling between $\mathcal{A}$ and $H_d$ is real: no new CP phase is introduced.}

\section{Cosmology of the axion multiplet}
\label{sec:cosmology}

In this section we discuss the cosmological issues of the axion multiplet.
The scalar component of the multiplet is composed of an axion, which obtains its mass from the QCD strong dynamics, and a saxion, which obtains the mass from the SUSY breaking through the K{\"a}hler potential. The fermion component of the multiplet is absorbed into the longitudinal component of the gravitino. We first consider the case where the cutoff scale of the SUSY breaking sector is equal to  the Planck scale ($M_* = \mpl$) and discuss several possible cosmological problems and observational constraints on the axion. We then consider the case with a (slightly) lower cutoff scale and argue that all of the problems are avoided.

The saxion in general has a large initial field value and a large energy density in the early universe. It is long-lived and may cause cosmological problems.
We first derive the condition such that the saxion decays before Big Bang nucleosynthesis (BBN). With the cutoff scale around the Planck scale, the coefficients of the K{\"a}hler potential are $O(1)$ in the Planck units. The saxion mass $m_S$ is then expected to be $O(m_{3/2})$.
The decay rate of the saxion becomes maximum when its decay into a pair of gravitinos is kinematically allowed. The decay rate is given by~\cite{Endo:2006zj}
\begin{eqnarray}
\Gamma(s \rightarrow 2\psi_{3/2}) = \frac{1}{96\pi} \frac{m_s^5}{m_{3/2}^2 \mpl^2}.
\end{eqnarray}
The gravitino decays into MSSM particles with a decay rate 
\begin{eqnarray}
\Gamma_{3/2} = \frac{121}{192\pi} \frac{m_{3/2}^3}{\mpl^2}.
\end{eqnarray}
For $m_{3/2}\gsim 100$ TeV, the saxion as well as the produced gravitino decays into MSSM particles before the onset of the BBN.
The gaugino mass of $O(1)$ TeV is obtained for $f_a \gsim 10^{18}$\,GeV.

Even if the saxion decays before the BBN, as its decay products eventually decay into the LSP, the universe may be overclosed by dark matter. We consider two solutions to the problem. 
We may simply assume that the saxion initial field value is fine-tuned. This may be required by the anthropic principle; a larger dark matter density leads to earlier collapse of dark matter into halos with larger densities, where habitable planets are more easily destroyed~\cite{Tegmark:2005dy}. 
Another possibility is that the $R$ parity is broken, and hence the LSP is unstable.
This may make the higgsino unstable in the collider time scale, so that the LHC can more easily search for the higgsino. For example, with the $R$ parity violating operator $\kappa L_{\tau} H_u$, the neutral higgsino decays into $\tau^{\pm} + W^\mp$, and the charged higgsino decays into $\tau +  Z/h$ if $\kappa$ is sufficiently large. The LHC with $\sqrt{s} =13$ GeV and the integrated luminosity of $1{\rm ab}^{-1}$ can search for the higgsino as heavy as $600$ GeV~\cite{Kumar:2015tna}.

The saxion also decays into axions through the interaction
\begin{eqnarray}
{\cal L} = \frac{1}{\sqrt{2}} \frac{\partial^3 K}{\partial x^3} s \partial a \partial a,
\end{eqnarray}
which are observed as dark radiation of the universe. Using Eqs.~(\ref{eq:vcc}) and (\ref{eq:vst}), the decay rate is given by
\begin{eqnarray}
\Gamma(s \rightarrow 2a) =  \frac{1}{48\pi} \frac{m_{s}^3}{\mpl^2}.
\end{eqnarray}
When the saxion dominates the energy density of the universe and decays, which is allowed if the $R$ parity is violated, the abundance of the axion is given by
\begin{eqnarray}
\Delta N_{\rm eff} = 0.34 \times \left( \frac{6 m_{3/2}}{m_s}\right)^2.
\end{eqnarray}
In order for the abundance of the dark radiation to satisfy the experimental constraint, $\Delta N_{\rm eff} < 0.3$~\cite{Planck:2018vyg}, the saxion mass is required to be about six times larger than the gravitino mass.
When the initial saxion abundance is fine-tuned, so that the LSP abundance is suppressed without the $R$ parity violation, the saxion is a subdominant component of the universe when it decays. The abundance of the axion produced from the decay of the saxion is negligibly small.

With the decay constant of $f_a \sim \mpl$, the axion abundance produced by the initial misalignment angle~\cite{Preskill:1982cy,Abbott:1982af,Dine:1982ah} exceeds the observed dark matter abundance, if the initial angle is larger than $O(10^{-4})$. We assume that the initial angle is fine-tuned to be small by the anthropic principle.
Note that the axion abundance much smaller than the observed one requires extra fine-tuning in the initial angle. Thus we expect that the axion abundance is comparable to the observed dark matter abundance. Axion dark matter with a decay constant of $O(\mpl)$ can be detected by proposed experiments~\cite{Budker:2013hfa}.

A QCD axion with a decay constant around the Planck scale has a Compton length comparable to the sizes of astrophysical black holes.
Such a bosonic particle, through the black hole-superradiance effect~\cite{Penrose:1969pc}, slows down the rotation of black holes. From the observations of black holes with large spins, the existence of light bosons might be excluded~\cite{Arvanitaki:2014wva}.
The determination of the spin of black holes, however, crucially depends on the modeling of accretion discs as well as emission of X-rays from them, which is currently subject to some uncertainties (see e.g., \cite{Middleton:2006kj,Middleton:2015osa,Kawano:2017lqd}).
We conservatively consider that the decay constant of the Planck scale is still a viable option.
Alternatively, we can lower $f_a$ to be $\sim 10^{17}$ GeV by introducing a (moderately) large number of charged particles under the shift symmetry as shown in Appendix~\ref{sec:app1}. In this case, the axion mass is large enough to avoid the above constraint. However, the $O(1)$ TeV gaugino masses require $m_{3/2} \sim 10$\,TeV, and the saxion decays after the onset of the BBN. The BBN constraint requires a saxion abundance much smaller than that required by the LSP overproduction.


We next consider the case with a cutoff scale of the SUSY breaking sector lower than the Planck scale. In this case, the saxion mass is above the gravitino mass. As is shown in Appendix~\ref{sec:app1}, $f_a$ is below the Planck scale. Then $O(1)$ TeV gaugino masses require $m_{3/2} < 100 $ TeV and the gravitino decays after the onset of the BBN. Since the gravitino is produced from the saxion decay, the saxion abundance must be small. This can be naturally achieved by the adiabatic suppression mechanism~\cite{Linde:1996cx, Nakayama:2011zy, Nakayama:2012mf} without the fine-tuning, owing to the small cutoff scale. Also, R parity violation is not required since the LSP overproduction from the saxion decay is simultaneously avoided. The saxion does not dominate the universe and hence the abundance of axions as dark radiation is negligible. Finally, because of $f_a < \mpl$, the superradiance constraint is avoided. The only possible prorblem is the axion overproduction by the initial misalignment angle, but that may be explained by the anthropic requirement as discussed above.

\section{Discussion and conclusions}
\label{sec:discussion}
In this paper we consider a simple theory where a single string axion multiplet is responsible for supersymmetry breaking, its mediation to the standard model sector, and a solution to the strong CP problem. The couplings of the axion multiplet to the gauge multiplets are fixed by the anomaly indices of the shift symmetry, and hence the gaugino masses take fixed, rational ratios. Assuming that the soft masses of scalars vanish at the mediation scale, focus point gaugino mediation is realized.
The electroweak scale is obtained by a tuning of only $O(1)$\% between the gaugino mass and the supersymmetric mass term of the Higgs multiplet.


We also discuss the cosmology of the axion multiplet. If the cutoff scale of the supersymmetry breaking sector is around the Planck scale, the BBN constraint is avoided for the axion decay constant and the gravitino mass around $O(\mpl)$ and $O(100)$ TeV, respectively. The overproduction of the LSP dark matter from the decay of the saxion is avoided by the anthropic principle or $R$ parity violation. For the former case the higgsino composes (a part of) dark matter, and signals in near future direct detection experiments are expected. For the latter case the axion produced by the decay of the saxion may be observed as an extra relativistic component of the universe. If the cutoff scale of the supersymmetry breaking sector is below the Planck scale, the BBN and LSP-overproduction constraints on the saxion abundance is naturally avoided by the adiabatic suppression mechanism.

With a decay constant not much below $O(\mpl)$, the axion abundance produced by the misalignment exceeds the observed dark matter abundance, if the misalignment angle is $O(1)$. Assuming the suppression of it by the anthropic principle, the axion is expected to compose $O(1)$ fraction of the dark matter density in the universe. The dark matter axion can be detected in proposed experiments.

\section*{Acknowledgement}
This work is supported by
the Office of High Energy Physics of the U.S. Department of Energy under Contract DE-AC02-05CH11231 (K.H.),
the National Science Foundation under grant PHY-1316783 (K.H.),
JSPS KAKENHI Grant Numbers JP26104009 (T.T.Y), JP26287039 (T.T.Y.), JP16H02176 (T.T.Y), JP15H05889 (N.Y.), JP15K21733 (N.Y.), JP17H05396 (N.Y.), JP17H02875 (N.Y.), 
and by World Premier International Research Center Initiative (WPI Initiative), MEXT, Japan (T.T.Y.).
K.H.~thanks the hospitality of Kavli IPMU, where a part of this work has been performed.
N. Y. is supported by a start-up grant from Zhejiang University. 
T. T. Y. is supported in part by the
China Grant for Talent Scientific Start-Up Project and by Natural Science Foundation of
China (NSFC) under grant No. 12175134.

\appendix
\section{Gravitational SUSY breaking with a lower cut-off and axion induced gaugino masses} \label{sec:app1}
In this appendix, we show that the cut-off scale for the gravitational SUSY breaking, $M_*$, can be lower than $\mpl$ as shown in Ref.~\cite{Asano:2015kvj}.  We also show a concrete setup generating the gaugino masses as well as a generation of $f_a \sim 10^{17}$\,GeV.  Let us consider the SUGRA Lagrangian 
\begin{eqnarray}
\mathcal{L} \ni \int d^4 \theta \Phi^\dag \Phi \left[-3 \mpl^2 e^{-K/(3\mpl^2)}\right] + \int d^2 \theta \Phi^3 W + h.c. ,
\end{eqnarray}
where $\Phi = \phi(1 + F_\Phi \theta^2)$ is the conformal compensator and the K{\"a}hler potential, $K$,  takes the sequestered form,
\begin{eqnarray}
K = -3 \mpl^2 \ln \left[1 - \frac{M_*^2 g(x)}{3 \mpl^2}  - \frac{f_{\rm vis}}{3 \mpl^2} \right].
\end{eqnarray}
Here $x=Z+Z^*$ and the Lagrangian is invariant under the shift transformation, $Z \to Z + i\, \mathcal{R}$  with $\mathcal{R}$ being a real constant,  $f_{\rm vis}$ is a function of chiral superfields in the visible sector, and $\phi$ is chosen to be 
$\phi=e^{\left<K\right>/(6\mpl^2)}$ so that the Einstein frame is realized.
The scalar potential for $Z$ is given by
\begin{eqnarray}
-V(x) &=& |\phi|^2 \left[ |F_\Phi|^2 f(x) + F_\Phi^* F_Z f'(x) + F_\Phi F_Z^* f'(x) + |F_Z|^2 f''(x) \right] \nonumber \\
&+& 3 \phi^3 F_\Phi W + 3 \phi^{*3} F_\Phi^* W^*,
\end{eqnarray}
where $f=-3 \mpl^2 e^{-K/(3\mpl^2)} = - 3 \mpl^2 + M_*^2 g(x)$. 
Using the equations of motions, the $F$-terms of $\Phi$ and $Z$ are 
\begin{eqnarray}
F_\Phi &=& \frac{3 f''\phi^{*3} W^*}{|\phi|^2(f'^2 -f f'') }, \nonumber \\
F_Z &=& -\frac{3 f' \phi^{*3} W^*}{|\phi|^2(f'^2 -f f'') },
\end{eqnarray}
and the scalar potential becomes
\begin{eqnarray}
V &=& -3 \phi^3 F_\Phi W.
\end{eqnarray}
Therefore, the vanishing cosmological constant is obtained for $F_\Phi =0 \to f''(x)=g''(x)=0$ at the minimum that satisfies $V'(x)=0$ and $V''(x)>0$. These minimization conditions are satisfied for $g^{(3)}(x)=0$ and $g^{(4)}(x)<0$. At the minimum,
\begin{eqnarray}
\left<F_Z\right> = -\frac{\mpl^2}{M_*^2} \frac{3 m_{3/2}}{|\phi|^2\left<g'(x)\right>}, \  
|\phi|^2 = \left( 
1 - \frac{M_*^2 \left<g(x)\right> }{3 \mpl^2}
\right)^{-1}.
\end{eqnarray}
Therefore, SUSY is broken.  We see that $\left<F_Z\right>$ can be quite largely enhanced by $\mpl^2/M_*^2$ in comparison with $m_{3/2}$.
If the shift symmetry is anomalous, $Z$ couples to field-strength superfields as
\begin{eqnarray}
\mathcal{L} \ni \frac{1}{32\pi^2} \int d^2 \theta Z (k'_1 \mathcal{W}_1^2
+k'_2 \mathcal{W}_2^2
+k'_3 \mathcal{W}_3^2) + h.c. \label{eq:anom_ww}
\end{eqnarray}
These terms can arise from interactions, for instance~\cite{Higaki:2011bz},  
\begin{eqnarray}
W = M' e^{-q_a Z} \Psi_{a}^{I_a} \bar{\Psi}_{a}^{I_a},
\end{eqnarray}
where $a$ represents a gauge index, $I_a=1 \dots N_a$ is the number of charged matters, and $\Psi_{a}^{I_a}$ are assumed to be fundamental representations of the gauge groups. $\Psi_{a}^{I_a} \bar{\Psi}_{a}^{I_a}$ transforms as $\Psi_{a}^{I_a} \bar{\Psi}_{a}^{I_a} \to \exp[i q_a \mathcal{R}] \Psi_{a}^{I_a} \bar{\Psi}_{a}^{I_a}$ under the shift symmetry.
In this case, $k'_a = q_a N_a$. The charge $q_a$ is assumed to be quantized.
The gaugino masses read~\footnote{
Sfermion masses can be suppressed in an extra-dimensional setup~\cite{Kaplan:1999ac} or by introducing copies of the gauge groups~\cite{Green:2010ww,Auzzi:2010mb,Sudano:2010vt}.
}
\begin{eqnarray}
M_a = -\frac{k'_a  g_a^2}{16\pi^2} \left<F_Z\right>.
\end{eqnarray}

To discuss the axion couplings, we define a canonically normalized field $\mathcal{A}$, 
\begin{eqnarray}
\left<K_{x x}\right>^{1/2} (Z-\left<Z\right>) = \mathcal{A},
\end{eqnarray}
where
\begin{eqnarray}
\left<K_{x x}\right> 
= \left<\frac{\partial^2 K}{\partial x^2}\right> = \frac{1}{3} \frac{M_*^4}{\mpl^2} \left< g'(x)^2 \right> 
\left(
1 - \frac{M_*^2\left< g(x) \right>  }{3 \mpl^2} 
\right)^{-2} \equiv \left( \frac{\mathcal{N}}{\sqrt{2}}  \frac{M_{*}^2}{\mpl} \right)^2.
\end{eqnarray}
Here, $\mathcal{N}$ is an $\mathcal{O}(1)$ constant. Note that, due to the requirement of the vanishing cosmological constant, $\left<g''(x)\right>=0$, the normalization factor $\left<K_{x x}\right>^{1/2} \sim M_*^2/M_{Pl}$ rather than $\sim M_*$.
This canonically normalized field, $\mathcal{A}$, has the $F$-term of $\left<F_{\mathcal{A}}\right>= - \sqrt{3} m_{3/2} \mpl$. The saxion mass, $m_s$, is 
\begin{eqnarray}
m_s = \sqrt{2 \left< \frac{\partial^2 V}{\partial x^2}/K_{xx} \right>} =   m_{3/2} \left(\frac{\mpl}{M_*}\right)^{3}
\sqrt{-54
\left<
\frac{g^{(4)}(x)}{(g'(x))^4 } \left(1- \frac{M_*^2 g(x)}{3\mpl^2}\right)^3 \right>}.
\end{eqnarray}
Equation (\ref{eq:anom_ww}) is written by
\begin{eqnarray}
\mathcal{L} &\ni& \frac{\sqrt{2}}{32\pi^2} \frac{\mpl}{\mathcal{N} M_*^2}  \int d^2 \theta \mathcal{A} (k'_1 \mathcal{W}_1^2
+k'_2 \mathcal{W}_2^2
+k'_3 \mathcal{W}_3^2) + h.c. \nonumber \\
&=& \frac{\sqrt{2}}{32\pi^2 f_a}  \int d^2 \theta \mathcal{A} (k_1 \mathcal{W}_1^2
+k_2 \mathcal{W}_2^2
+ k_3 \mathcal{W}_3^2) + h.c.,
\end{eqnarray}
where
\begin{eqnarray}
f_a = \frac{\mathcal{N} M_*^2}{k'_3 \mpl}, \ 
k_1 = k'_1/k'_3, \ 
k_2 = k'_2/k'_3, \ 
k_3 = 1.
\end{eqnarray}
Note that $f_a \sim 10^{17}$ GeV can be obtained by taking slightly smaller $M_*$ than $\mpl$ and/or large $k'_3$.

\end{document}